# The Structure of Martian Magnetosphere at the Dayside Terminator Region as Observed on MAVEN Spacecraft

**O. L. Vaisberg[1], V. N. Ermakov[1,2], S. D. Shuvalov[1], L. M. Zelenyi[1], J. Halekas[3], G. A. DiBraccio[4], J. McFadden[5], and E. M. Dubinin[6]**

[1] Space Research Institute of the Russian Academy of Sciences, Moscow, Russia.

[2] National Research Nuclear University MEPhI, Moscow, Russia.

[3] Department of Physics and Astronomy, University of Iowa, Iowa City, Iowa, USA.

[4] NASA Goddard Space Flight Center, Maryland, USA.

[5] Space Sciences Laboratory, U.C., Berkeley, Berkeley, CA, USA.

[6] Max-Planck-Institute for Solar System Research, Gottingen, Germany.

Corresponding author: Oleg Vaisberg (olegv@iki.rssi.ru)

**Key Points:**

- The dayside magnetosphere of Mars at ~70$^0$ SZA is a permanent domain of ~200 km thickness between the magnetosheath and the ionosphere

- Magnetopause is defined by a steep gradient of $O^+$ and $O_2^+$ densities which increase by factor of $10^2$-$10^3$ at interface with the ionosphere

- The structure of the magnetosphere is controlled by the solar wind magnetic field direction.

**Abstract**

We analyzed 44 passes of the MAVEN spacecraft through the magnetosphere, arranged by the angle between electric field vector and the projection of spacecraft position radius vector in the YZ plane in MSE coordinate system ($\theta_E$). All passes were divided into 3 angular sectors near 0°, 90° and 180° $\theta_E$ angles in order to estimate the role of IMF direction in plasma and magnetic properties of dayside Martian magnetosphere. The time interval chosen was from January 17 through February 4, 2016 when MAVEN was crossing the dayside magnetosphere at SZA ~ 70°. Magnetosphere as the region with prevailing energetic planetary ions is always found between the magnetosheath and the ionosphere. 3 angular sectors of dayside interaction region in MSE coordinate system with different orientation of the solar wind electric field vector $\boldsymbol{E}$ = -1/c $\boldsymbol{V}$x$\boldsymbol{B}$ showed that for each sector one can find specific profiles of the magnetosheath, the magnetic barrier and the magnetosphere. Plume ions originate in the northern MSE sector where motion electric field is directed from the planet. This electric field ejects magnetospheric ions leading to dilution of magnetospheric heavy ions population, and this effect is seen in some magnetospheric profiles. Magnetic barrier forms in front of the magnetosphere, and relative magnetic field magnitudes in these two domains vary. The average height of the boundary with ionosphere is ~530 km and the average height of the magnetopause is ~730 km. We discuss the implications of the observed magnetosphere structure to the planetary ions loss mechanism.

**Plain Language Summary**

As Mars does not have an intrinsic global magnetic field the solar wind directly interacts with the gaseous envelope of Mars. This interaction leads to formation of the magnetosphere from magnetic field tubes of the solar wind that bend around the planet forming magneto-plasma envelope around it. The dayside of the magnetosphere was not studied in detail due to its relatively small scale. MAVEN spacecraft with its comprehensive payload gives possibility for studying the dayside magnetosphere of Mars.

Analysis of MAVEN plasma and magnetic measurements showed that the dayside Martian magnetosphere is a permanent layer of the magnetized plasma between heated solar wind plasma flow and the ionosphere. With average thickness of ~200 km it is filled with planetary ions accumulated during convection of these tubes from the dayside to the tail. These ions then escape through the tail being the one of primary loss sources led to devastating the Martian atmosphere through millennia. It is found that the magnetic structure and planetary ion flux in dayside magnetosphere are asymmetric what is determined by the direction of the interplanetary magnetic field. This asymmetry is analyzed in the paper.

**1 Introduction**

      The obstacle to the solar wind flow at Mars has been observed for more than 45 years by several spacecraft. First crossings of the dayside magnetosphere with observations of increasing magnetic field up to ~30 nT (Dolginov et al., 1972) and ions of lower energy than magnetosheath ions were made by Mars-2 and Mars-3 (called then "ion cushion", Bogdanov and Vaisberg, 1975). The ion and magnetic tail at ~7 $R_M$ with planetary ions outflow was found on Mars-5 (Vaisberg et al., 1975b, Dolginov, 1978). The magnetic field increase, solar wind proton

depression and the presence of planetary ions were interpreted in terms of the boundary layer existence on the dayside (Szego et al, 1998).

Ion composition within the magnetic barrier at dayside of Mars was measured on Mars-Express (Dubinin et al., 2008, a, b). The cases of strong mass loading and the magnetic field pileup with discontinuity (the magnetic pileup boundary) were found as well as the crossings of the Martian magnetosphere on the dayside without a signature of a magnetic field pileup. The altitude of the magnetic barrier near subsolar point was 450-550 km independent on strong solar wind ram pressure variations. Rather sharp ionospheric boundary (PEB) with abruptly increasing photoelectron number density on this boundary up to $\sim 10^3$ cm$^{-3}$ was found.

Halekas et al. (2017a) stated that "the magnetosphere of Mars forms as a result of the direct and indirect interaction of the solar wind with the Martian ionosphere, through a combination of induction effects and mass loading". The Martian magnetosphere is dominated by plasma of atmospheric origin, which forms the primary global obstacle to the solar wind through induction and mass loading, with additional contributions from localized crustal magnetic fields (Halekas et al. 2017b).

Matsunaga et al. (2017) analyzed 10 months of MAVEN observations to determine the average locations of Ion Composition Boundary (ICB), Induced Magnetosphere Boundary (IMB) and Pressure Balance Boundary (PBB), They found that IMB almost coincide with ICB on the dayside.

Vaisberg et al. (2017) analyzed one case of Martian magnetosphere crossing by MAVEN spacecraft at terminator region, characterized by high mass loading. In this case the equality of magnetic and kinetic energies was observed which means that the regime of magnetospheric plasma flow is Alfvenic.

The structure and properties of the dayside Martian magnetosphere is much less studied than the night side of it due to its small scale and insufficient temporal resolution of previous Martian satellites. These may be the reasons for misconceptions of Marian magnetosphere including its name as "induced magnetosphere" and the conclusion that the solar wind directly interacts with Martian ionosphere.

Mars Atmosphere and Volatile EvolutioN mission (MAVEN) with comprehensive high-time resolution instruments suite provided an excellent possibility to study Martian environment. We analyze the structure and the properties of the dayside Martian magnetosphere near terminator in MSE coordinate system that shows the role of IMF direction in plasma and the magnetic properties of the dayside Martian magnetosphere.

The plan of the paper is the following. An example of MAVEN dayside pass at Mars is given to discuss how the boundaries of magnetosphere are identified. Selected passes through the magnetosphere were divided in 3 sectors depending on $\theta_E$ coordinate system. The typical properties of the magnetic barrier and the magnetosphere in each sector are summarized and exemplified. The dependence of the lower and the upper boundaries of magnetosphere in MSE coordinates is shown. The estimation of the heavy ion flux within the magnetosphere is given. The discussion of some magnetospheric properties is given.

## 2 Observations

We are using the data of following MAVEN instruments: STATIC (McFadden, 2015), SWIA (Halekas, 2017a), and MAG (Connerney et al., 2015, a, b).

The inward passes of MAVEN through dayside of solar wind-Mars interaction region were chosen in order to minimize the time interval between the pass of MAVEN from the solar wind to the magnetosphere. The time interval chosen was from January 17 through February 4, 2016 when MAVEN crossed the dayside magnetosphere at SZA ~70°. This time interval was chosen such that crossings of dayside magnetosphere were in northern hemisphere of Mars where the possible influence of magnetic anomalies is minimal. The flank region of magnetosphere was chosen for analysis so that it should be relatively far from the subsolar region where Martian magnetosphere is forming, and relatively close to terminator that provides plasma for the nightside magnetosphere of Mars.

The passes of MAVEN within selected time interval were arranged according to the $\theta_E$ angle (the angle between electric field vector and the projection of spacecraft position radius vector in the YZ plane in MSE coordinate system) in MSE coordinate system with X axis directed to the Sun, Z axis directed along vector $\boldsymbol{X} \times \boldsymbol{B_{SW}}$, and Y completing to the right system between the solar wind motional electric field vector $\boldsymbol{E} = -1/c\ \boldsymbol{V} \times \boldsymbol{B}$ direction. This coordinate system is also called the magnetic coordinate system (Figure 1). The solar wind electric field was computed from the table compiled by one of co-authors (SWIA PI J.Halekas) from SWIA and MAG (J.Connerney PI) measurements.

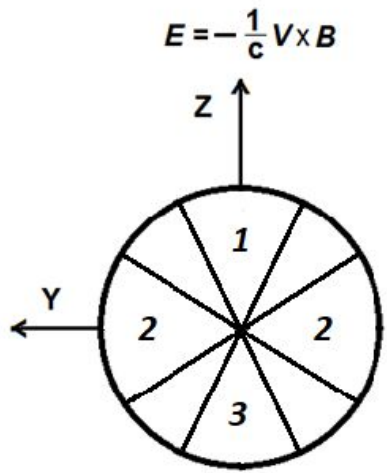

Figure 1. $\theta_E$ angular sectors in MSE coordinate system selected for analysis

Altogether 44 inbound passes were selected for analysis according to their positions in three angular intervals:

#1      0°-30° (18 crossings);

#2      60°-120° (14 crossings);

#3      150°-180° (13 crossings).

There was no allowance for sign of $\theta_E$ angle in this selection. The distribution of magnetosphere passes over calculated MSE $\theta_E$ angles is shown in Fig. 2.

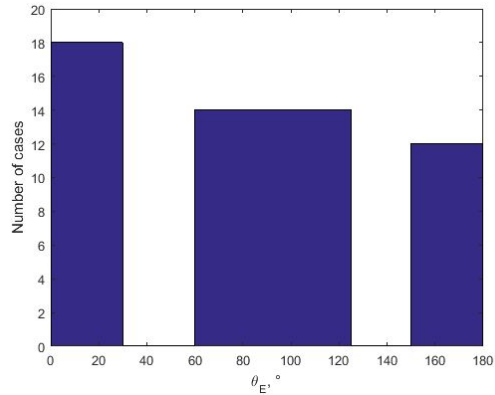

Figure 2. Distribution of passes through magnetosphere according to the $\theta_E$ angle in MSE coordinates. The height of each bar is the number within angular sector that is the width of the bar.

## 3    Analysis

### 3.1 An example

Initial analysis was made with the use of quick-looks plotted in AMDA database (a versatile web tool for Space Physics at the French national data centre for natural plasmas of the solar system, http://amda.cdpp.eu/) and quantitative plots with use of MatLab programs. Figure 3 shows the crossing of dayside solar wind-Mars interaction region from the group 1 ($\theta_E$ angle 13.35°). From left to right: the solar wind, the magnetosheath, the magnetosphere and the ionosphere.

The bow shock was crossed at ~ 09:19 UT as seen by the magnetic field jump and ion deceleration and heating. The magnetosheath continues until protons number density decrease and oxygen ions density increases at~ 09:43 UT. Noticeable oxygen ions are seen in the magnetosheath and they can be seen as more energetic (~3 keV) component on STATIC and SWEA energy-time spectrograms (the so called plume – oxygen ions ejected by the solar wind motional electric field from the magnetosphere-ionosphere (Liemohn et al., 2014, Dong et al., 2015, 2017, Andrews et al., 2017). The magnetic barrier (the steep monotonic rise of the magnetic field magnitude) was entered at ~ 09:32 UT.  The magnetosphere can be seen as oxygen-dominated region from magnetopause at ~ 09:43 UT till ionosphere at ~ 09:47:00 UT.

Magnetic field in the magnetic barrier continues to rise with maximum of ~23 nT in the magnetosphere, and then decreases to ~ 16 nT at the ionopause.

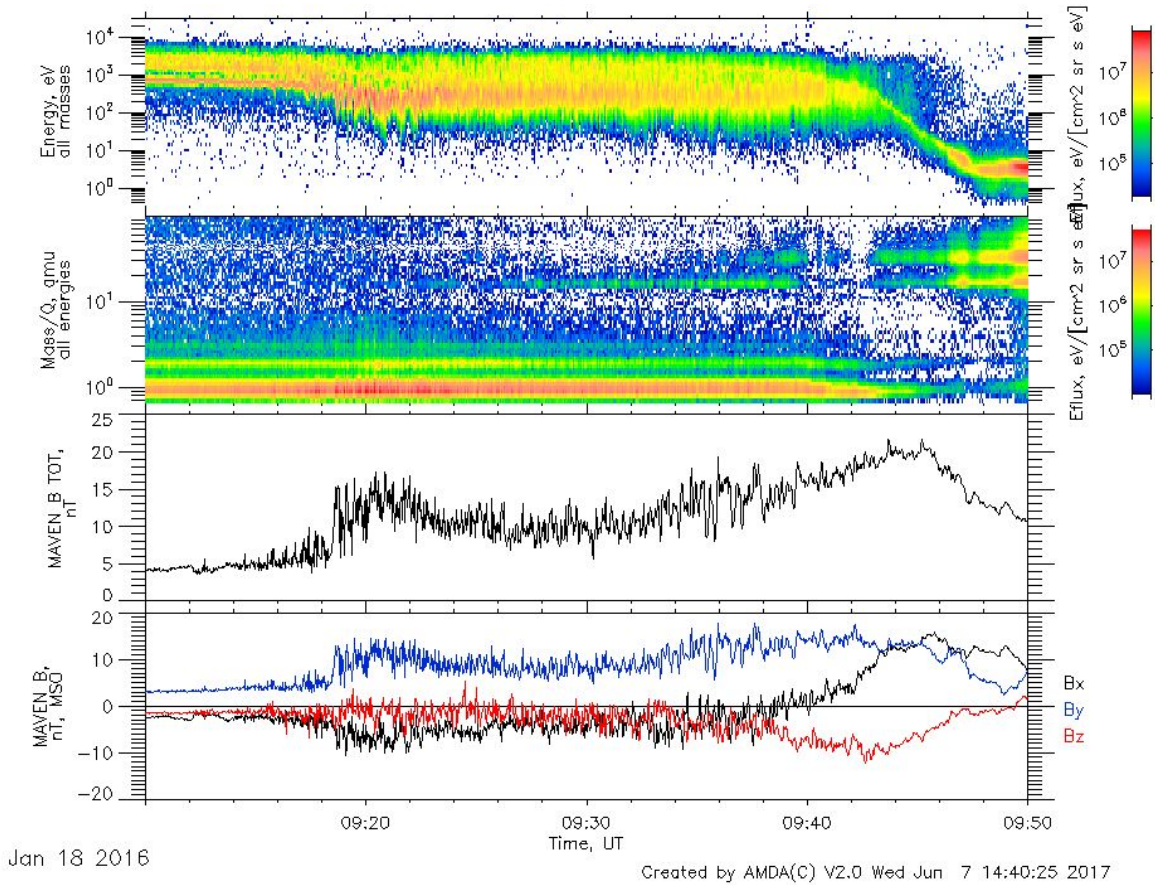

Figure 3. Overview of MAVEN inbound pass on January 18, 2016. From top to bottom: STATIC energy-time spectrogram (all ions), STATIC ion mass-time spectrogram, MAG magnetic field magnitude, MAG magnetic field components. AMDA (http://amda.cdpp.eu/). $\theta_E$ angle is 13.35°.

Figure 4 shows the properties of the plasma and the magnetic field in more detail. Numerical parameters allow us to determine the location of boundaries and identification of regions more precisely. The magnetosheath is usually defined by the domination of plasma flow energy. However, the magnetic barrier brings some correction due to the increase of magnetic energy and leads to partial loss of protons and alpha particles. The approximate equality of the plasma and the magnetic pressure characterizes the Alfvenic flow.

The magnetosphere in case shown is the region where the magnetic energy dominates. The heavy ions number density increases by more than 2 orders of magnitude from the magnetopause to the ionopause. The angle between the heavy ions velocity and the magnetic field direction is close to 90° that suggests an effective pick-up. The proton component in the magnetosphere moves at the angle ~140°-180° to the magnetic field direction that facilitates the loss of protons along the field line. Indeed, the loss of more energetic protons is seen in the energy-time spectrogram upstream of the magnetopause.

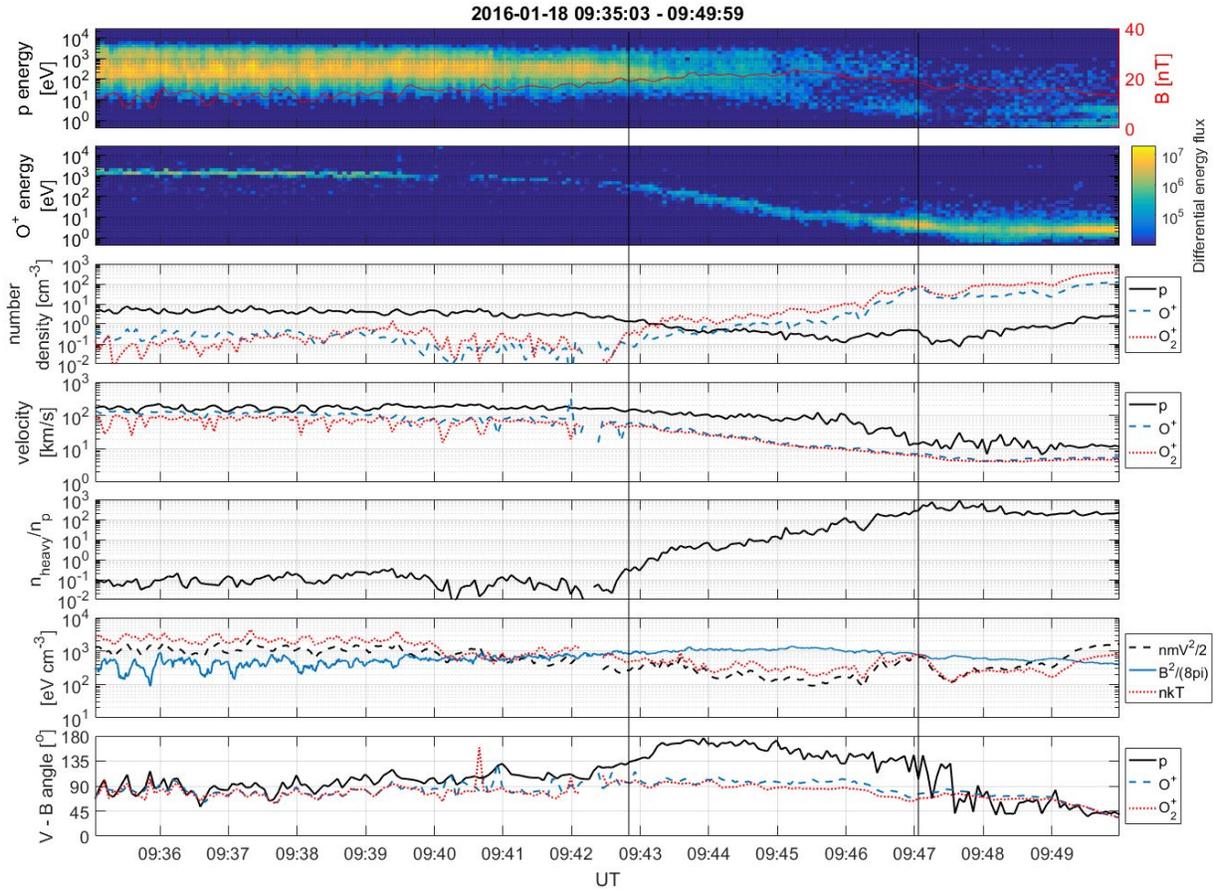

Figure 4. From top to bottom: the proton and O$^+$ energy-time spectrograms; the number densities of protons, O$^+$ and O$_2^+$ ions; the ratio of (O$^+$ + O$_2^+$) sum number density to proton number density; magnetic, kinetic, and thermal energy densities; the angles between the magnetic field direction and the ions ram velocities. The vertical lines define the magnetospheric boundaries. $\theta_E$ angle is 13.35°.

Considering magnetosphere as the region between the magnetosheath flow of the solar wind plasma and the ionosphere, we determined outer magnetosphere boundary (magnetopause) by n(O$^+$ + O$_2^+$)/n(p) ratio being in the range of 0.1 to 1 and/or its sharp increase in conjunction with proton energy drop. In most cases this boundary coincided with moderate increase of the magnetic field. Thus, the external boundary of the magnetosphere were similar to Ion Composition Boundary (ICB). It is worth noting, that in the sector of 0°-30° $\theta_E$ angle the magnetosphere location was complicated by plume heavy ions, which could disturb the magnetosheath and magnetospheric flows.

In order to determine the magnetosphere-ionosphere boundary, we used the heavy ions energy-time spectrograms. This boundary was identified as a transition from wider energy component with higher average energy to more dense, steady low-energy and cold heavy ions components (ionosphere). For 0°-30° sector, mostly observed with plume, this boundary was identified as a transition from energy component with higher average energy to steady low-energy and cold heavy ions components (ionosphere).

### 3.2 Sector 1: 0°-30°

Figures 5 and 6 show the typical magnetosphere crossings at high magnetic latitude, i.e. at small $\theta_E$ angle in MSE coordinate system.

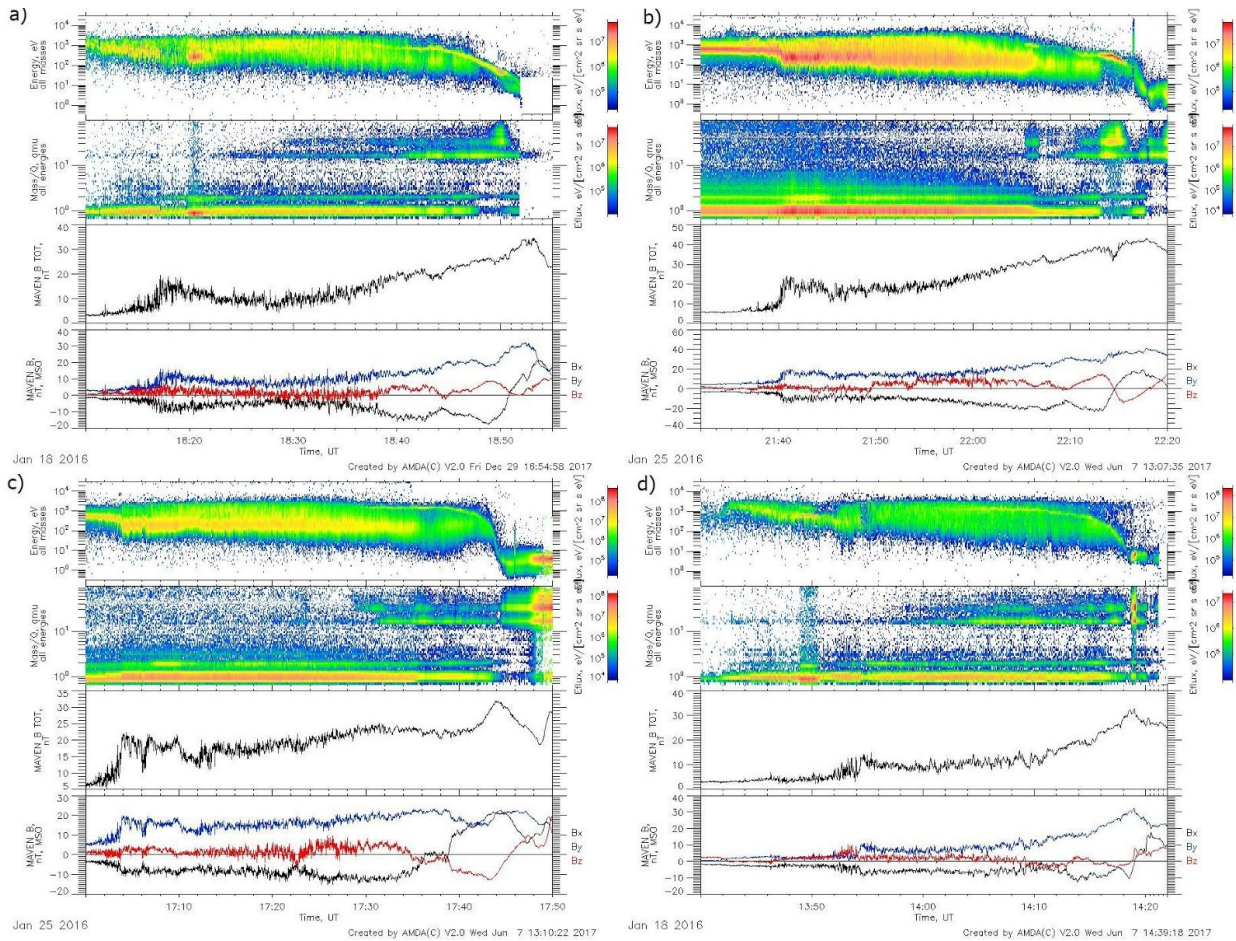

Figure 5. Four crossings of the magnetosphere at low $\theta_E$ angles (MSE sector #1). From top to bottom: energy-time spectrogram of all ions as measured by STATIC, the magnetic field magnitude and the three components measured by MAG. $\theta_E$ angles are: 7.26° (a), 23.11° (b), -12.27° (c), -11.42° (d).

Most of the crossings at high MSE latitude have many similar properties. They are:

1. Magnetosphere is always observed between the magnetosheath flow and ionosphere as a region dominated by planetary ions $O^+$ and $O_2^+$ with continuously decreasing velocity. The density of planetary ions increases by the factor of $10^2$ -$10^3$ from the magnetosheath to ionosphere.

2. Magnetopause as the outer boundary of magnetosphere is characterized by relatively sharp increase of $(n(O^+)+n(O_2^+))/n(p)$ ratio. In some cases heavy ions density at magnetopause is accompanied by the increase of magnetic field magnitude.

3. Magnetic barrier is found in all passes. It continues in magnetosphere with increasing module. Maximum B is usually observed within magnetosphere or at the boundary between the magnetosphere and the ionosphere. In most cases magnetic field magnitude is smaller in the ionosphere than in the magnetosphere.

4. Plume ions are seen in magnetosheath or magnetosphere or both. Strongest plume ions flux leads to the decrease of magnetic field magnitude in magnetic barrier in the magnetosphere.

5. Magnetic pressure dominates in the magnetosphere.

6. The density of magnetosheath protons decreases in the ionosheath/magnetic barrier layer adjacent to the magnetosphere. The density of protons strongly diminishes in the magnetosphere. The protons in the magnetosphere are moving at a relatively small angle to the magnetic field direction. The heavy ions move nearly perpendicular to the magnetic field direction.

7. The inner boundary of magnetosphere with ionosphere is difficult to locate due to existence of heated and accelerated ionospheric ions.

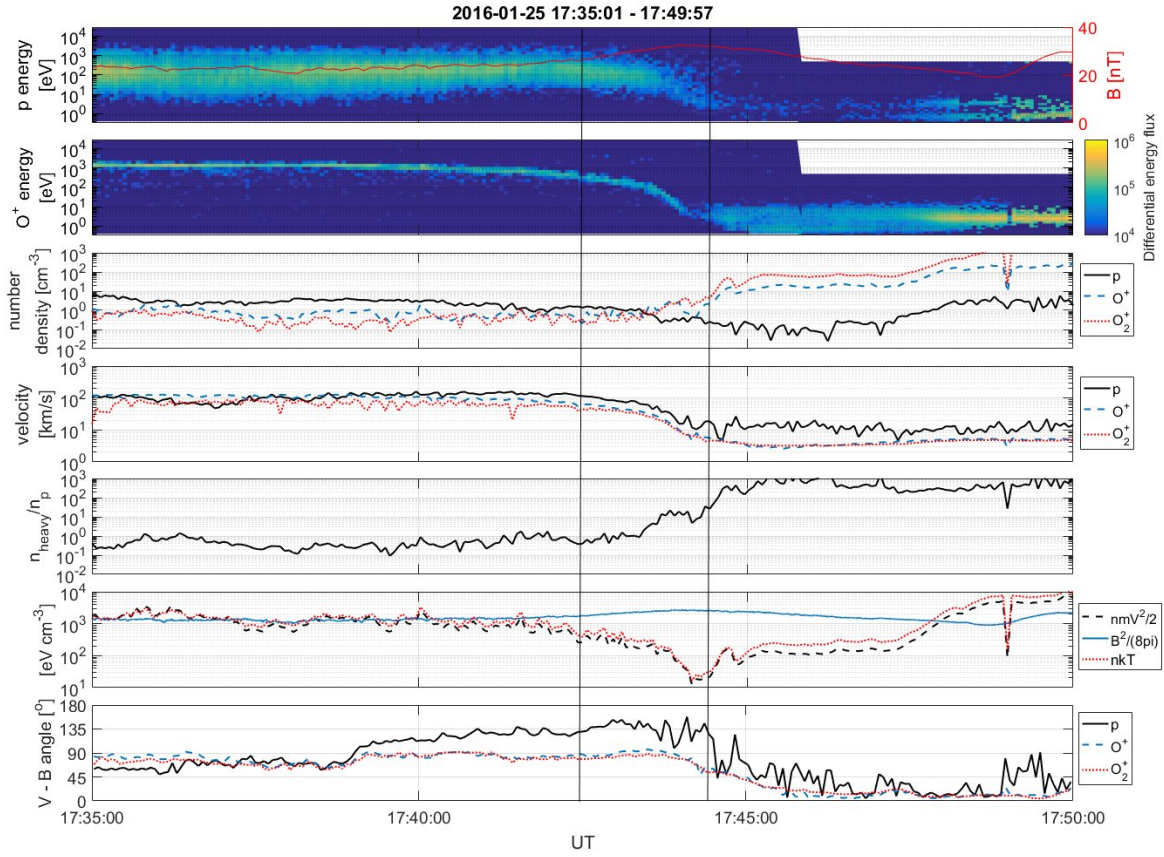

Figure 6. From top to bottom: proton and $O^+$ energy-time spectrograms; the number densities of protons, $O^+$ and $O_2^+$ ions; the ratio of $(O^+ + O_2^+)$ number density to proton number density; magnetic, kinetic, and thermal energy densities; the angles between the magnetic field direction and the ions ram velocities. The vertical lines define magnetospheric boundaries. Note the dominance of the magnetic energy in the magnetosphere and in the upper part of the ionosphere. $\theta_E$ angle is -12.27°.

### 3.3 Sector 2: 60°-120°

The structures of the magnetosheath, the magnetic barrier and the magnetosphere in this region of MSE are very variable. Figures 7 and 8 show four crossings of magnetosphere in this sector.

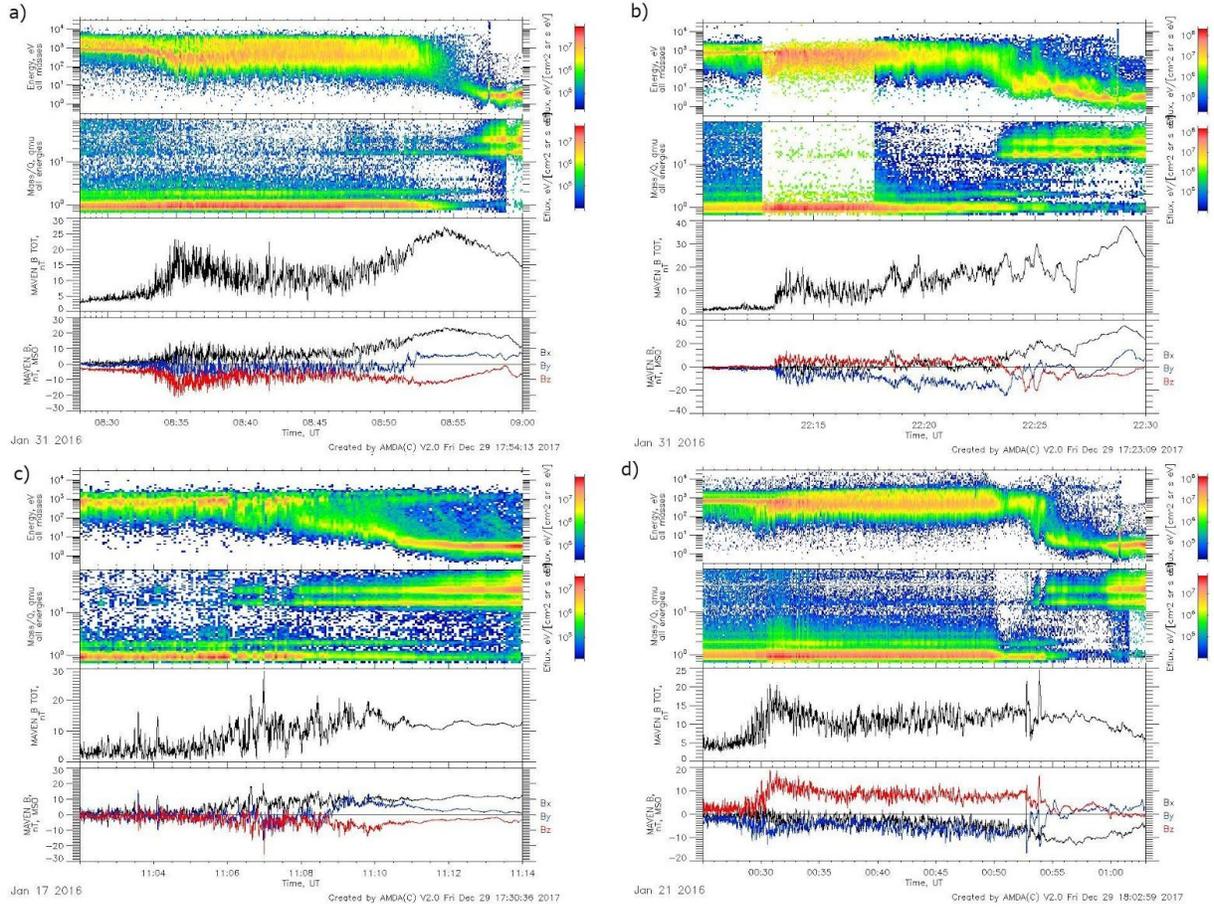

Figure 7. Four crossings of the magnetosphere at about 90° $\theta_E$ angles (MSE sector #2). From top to bottom: energy-time spectrogram of all ions as measured by STATIC, the magnetic field magnitude and the three components measured by MAG. $\theta_E$ angles are: -122.79° (a), -123.66° (b), -62.07° (c), 106.92° (d).

The following properties of the magnetosphere and its surrounding in this region are:

1. Magnetic barrier and magnetosphere profiles are quite variable in this sector
2. Magnetic barrier is frequently structured and may be less developed.
3. Magnetosheath proton flux is diluted in front of magnetosphere, density of higher energy protons decreases.
4. Magnetic field magnitude has maximum in the magnetosphere.
5. There are cases of multiple crossings of magnetopause
6. The structure of plasma in magnetosphere is disturbed.
7. There are energy dispersed signatures in the magnetosphere.

Figures 8a and 8b show the properties of the plasma, the magnetic field and the derived parameters in two in more detail. The pass on 01.31.2016 at ~ 08:44-09:02 UT shows the well ordered magnetosphere in which magnetic field pressure dominates. The pass on 01.31.2016 at ~ 22.21- 09:02 UT shows very chaotic structure. Plasma and magnetic pressures are variable with changing ratio of magnitudes.

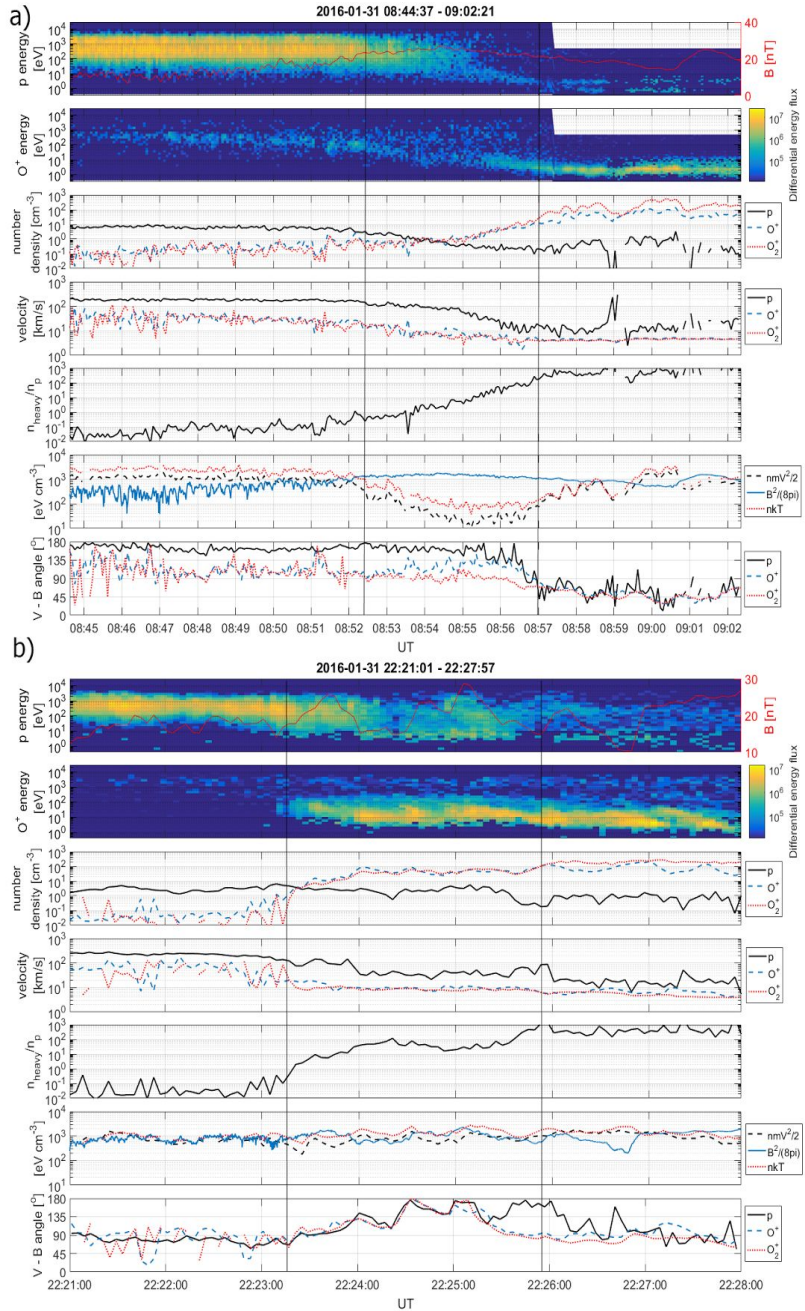

Figure 8. From top to bottom: proton and O$^+$ energy-time spectrograms; number densities of protons, O$^+$ and O$_2^+$ ions; the ratio of sum (O$^+$ + O$_2^+$) number density to proton number density;

magnetic, kinetic, and thermal energy densities; the angles between the magnetic field direction and the ions ram velocities. $\theta_E$ angles are: -122.79° (a), -123.66° (b).

### 3.4 Sector 3: 160°-180°

Figure 9 shows 4 examples of MAVEN passes in 160°-180° MSE sector.

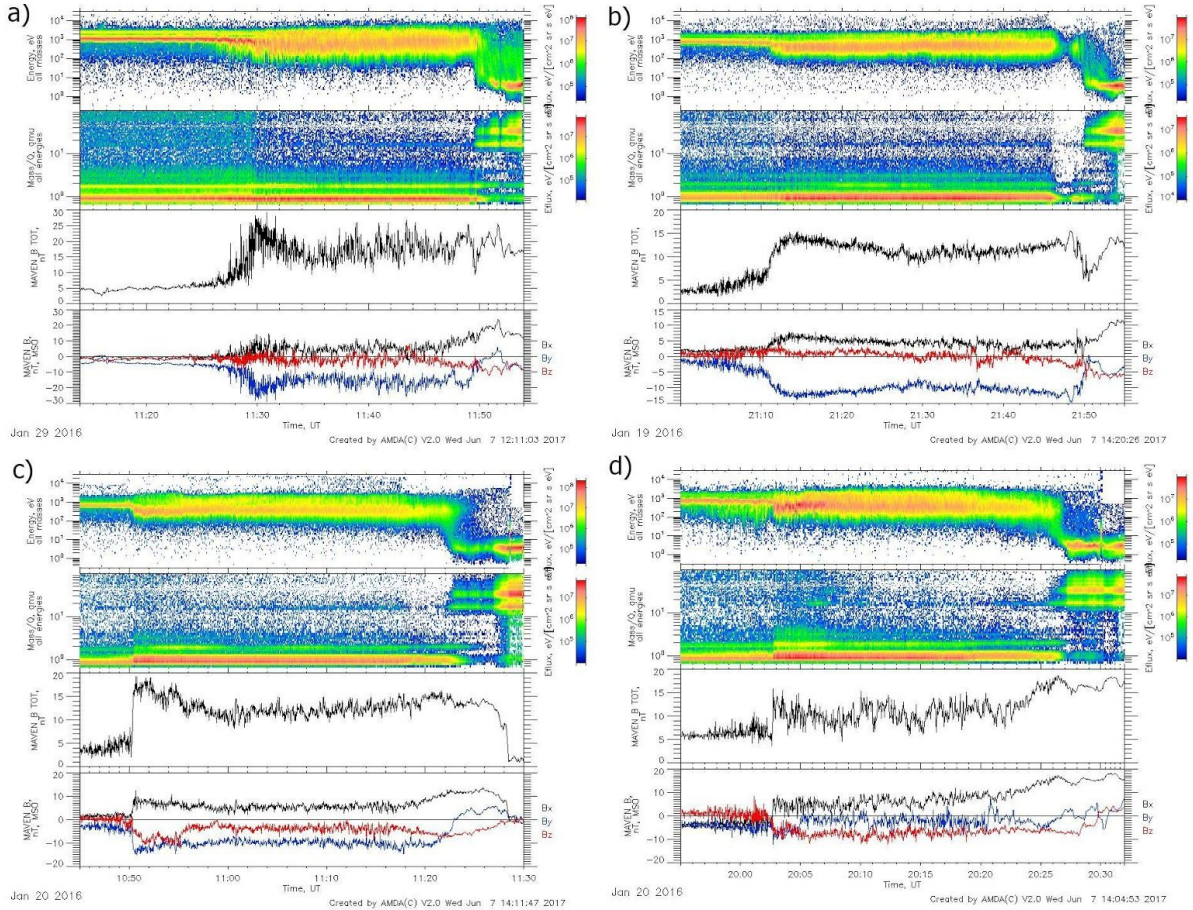

Figure 9. Four examples of MAVEN passes at 160°-180° $\theta_E$ angles (MSE sector #3). From top to bottom: energy-time spectrogram of all ions as measured by STATIC, the magnetic field magnitude and the three components measured by MAG. $\theta_E$ angles are: -179.29° (a), 167.53° (b),-168.82° (c), -161.98° (d).

Most of the crossings at high MSE latitude have many similar properties. They are:

1. Mass loading of magnetosheath is small and consists of $O^+$ ions
2. Magnetic field in magnetic barrier is weak and quite thin
3. The proton velocity in the magnetosheath is nearly constant till the magnetopause.
4. There is a region of smaller protons number density in front of the magnetosphere.
5. Magnetic field often has minimum at magnetopause.

6. Density gradient of $O^+$ and $O_2^+$ ions at magnetopause is higher at magnetopause than in other sectors.
7. The magnetic field magnitude is sometimes smaller in the magnetosphere than in the adjacent magnetosheath.
8. Magnetospheric ions have relatively small energy and sometime wide energy spectrum
9. The thickness of the magnetosphere is small.

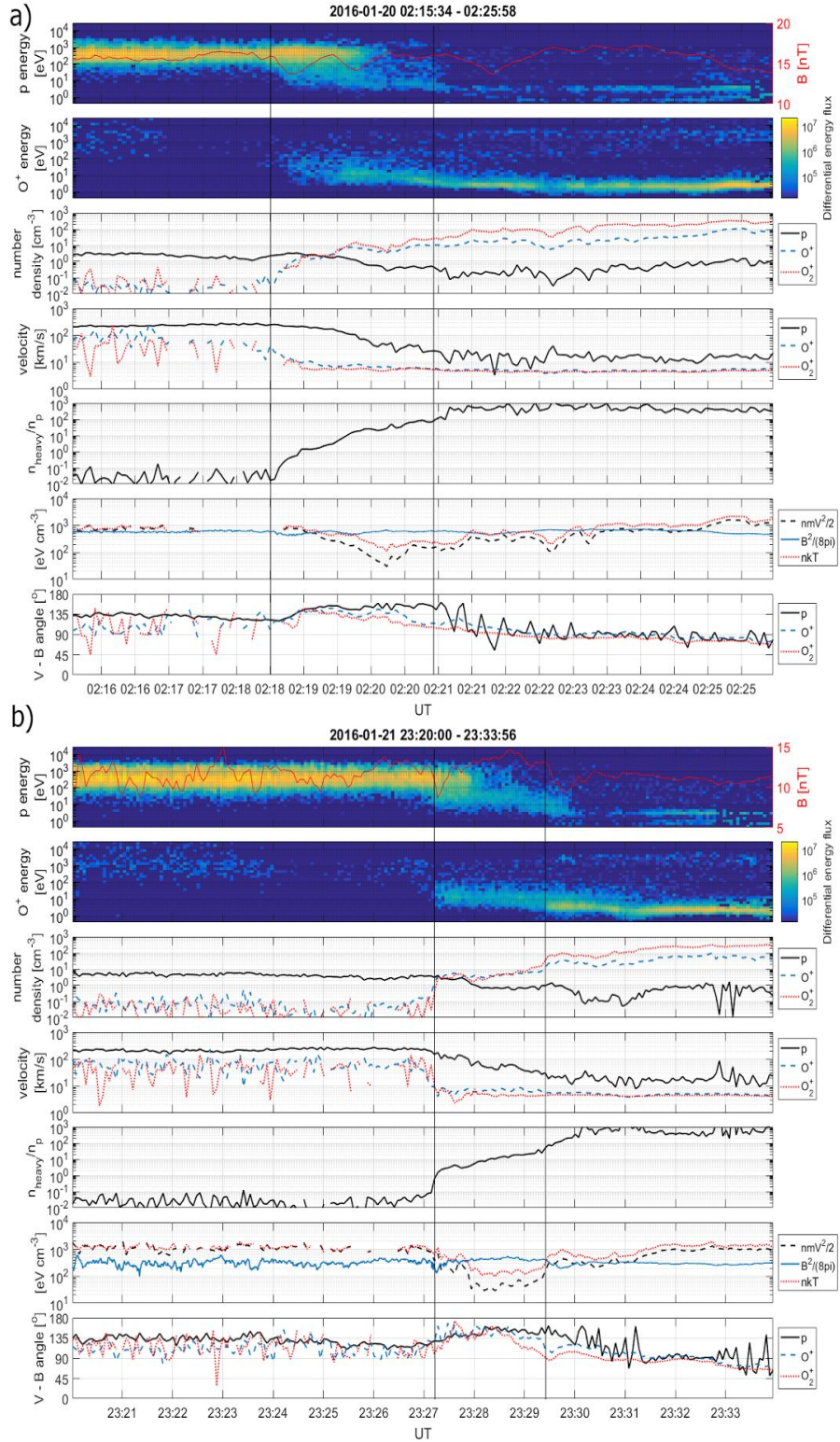

Figure 10. From top to bottom: proton and O$^+$ energy-time spectrograms; number densities of protons, O$^+$ and O$_2^+$ ions; the ratio of (O$^+$ + O$_2^+$) sum number density to proton number density;

magnetic, kinetic, and thermal energy densities; the angles between magnetic field direction and the ions ram velocities. $\theta_E$ angles are: -179.34° (a), -165.54° (b).

The two passes across the magnetosphere in the sector are shown in Fig. 10a and 10b: on 01:20:2016 at ~ 02:20 UT and 01:21:2016 at ~ 23:28 UT. One can see that energy distributions are different in this $\theta_E$ sector.

### 3.5 Boundaries

The distributions of the lower boundary of the magnetosphere height and the magnetopause height in three magnetic latitude intervals are shown in Figure 11.

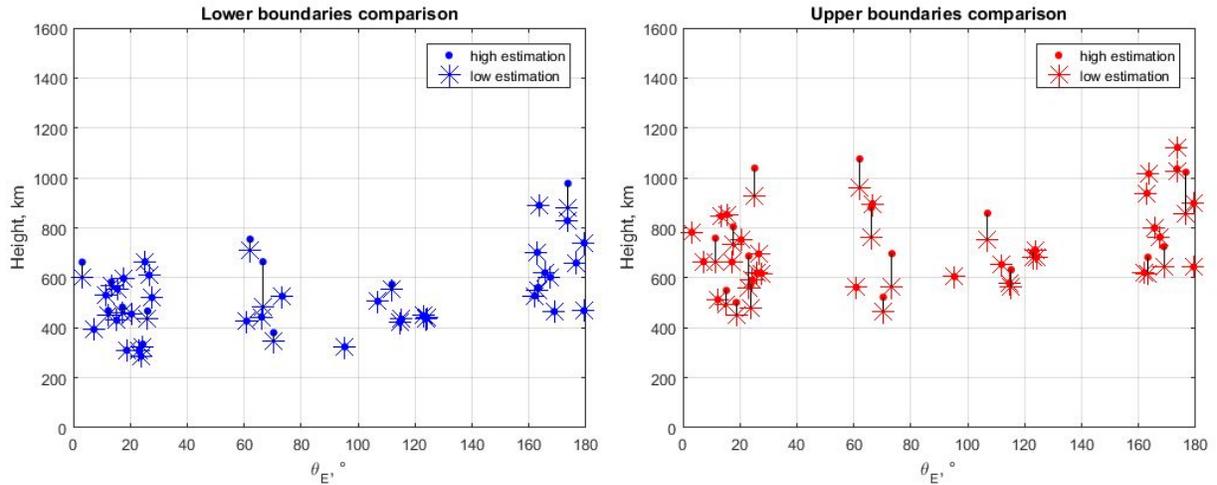

Figure 11. Height distributions of the magnetosphere boundaries in selected magnetic latitudes (MSE coordinates) intervals. Left: the boundary between the magnetosphere and the ionosphere, right - the magnetopause.

The heights of the boundaries determined as mentioned above and magnetosphere thicknesses are shown in Table 1. For the passes where authors were uncertain about the location of the boundaries, both the lower and the upper estimations were accounted as separate measurements.

Table 1. Magnetosphere boundaries heights and magnetosphere thickness at different MSE latitudes

| Sector | Magnetopause height, km | Magnetosphere-ionosphere boundary height, km | Magnetosphere thickness, km |
|---|---|---|---|
| 0°-30° | 673 ± 30 | 479 ± 23 | 200 ± 16 |
| 60° -120° | 705 ± 34 | 493 ± 28 | 223 ± 22 |

| 160° - 180° | 838 ± 43 | 661 ± 45 | 183 ± 17 |

### 3.6 Heavy ions flux

The approximate flux of heavy ions through the vertical cut of the magnetosphere with the width of 1 cm was calculated by the height-integration of the measured flux along the trajectory of the spacecraft across magnetosphere:

$$f_i = \int_{h_{min}}^{h_{max}} n_{b_i} v_{b_i} dh \,,$$

where $i$ stands for the number of ion mass, $f_i$ is the flux of this ion mass, $n_b$ and $v_b$ are the number density and the bulk velocity, respectively, $h$ is the height, $h_{min}$ and $h_{max}$ are the boundaries of the magnetosphere. The fluxes were calculated separately for three ions: p, $O^+$, and $O_2^+$ for each pass through the magnetosphere (Figure 12). As the field of view of STATIC frequently does not cover all sphere, each value needs to be considered as the lower value. The ion flux within sector # 1 of the magnetosphere is often seen as a plume, its flux within the magnetosphere was included in the calculated flux. The median value of the calculated fluxes of heavy ions is about f ~ 6x10$^{13}$ cm$^{-1}$ s$^{-1}$. Multiplying this value by the length of the circle of magnetosphere at SZA ~70° we receive the estimation of integral heavy ions flux $f_{int}$ through the magnetospheric cross section $f_{int}$ = f*2$\pi$R$_M$~1.3*10$^{23}$ s$^{-1}$. This is lower value as STATIC field of view does not cover all sphere and part of ion velocity distribution is not measured.

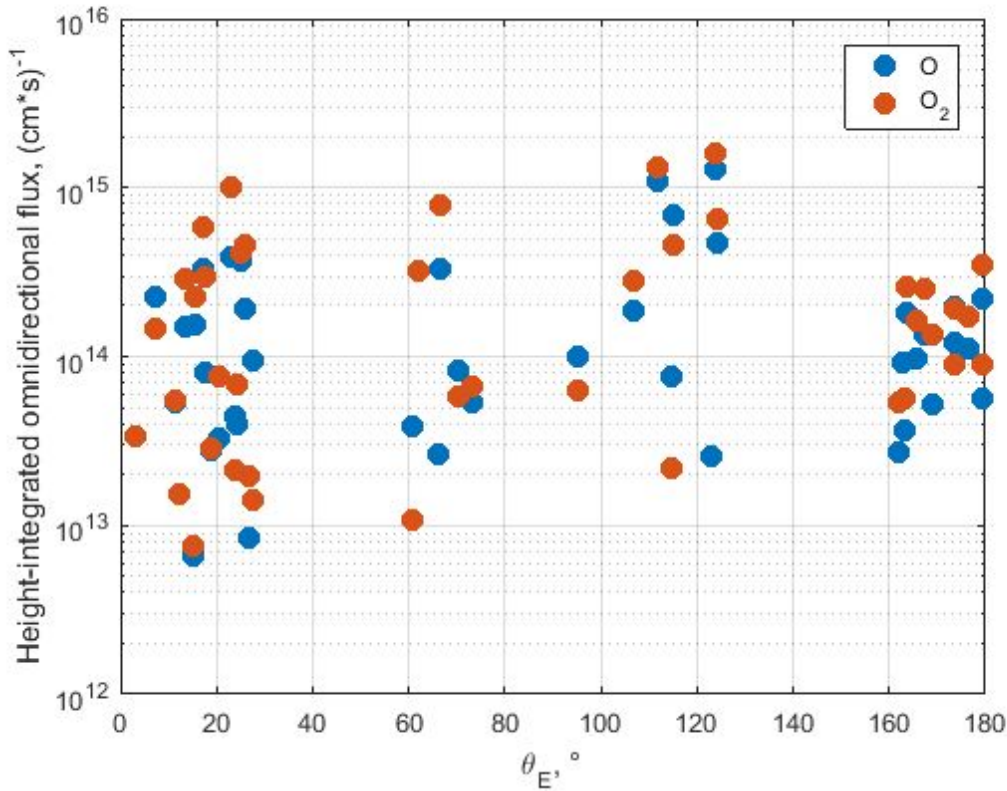

Figure 12. The calculated values of flux densities through 1 cm vertical cut of the magnetosphere.

**4 Discussion**

All 44 MAVEN passes on Martian dayside at ~ 70° SZA angle considered in this paper showed the existence of the region between the magnetosheath flow and the ionosphere that we consider as the magnetosphere. In spite of relatively small thickness, about 200 km at SZA ~ 70°, it plays important role in the formation of the nightside magnetosphere and the source of ions transported to the nightside and apparently lost through the tail. The properties of magnetosphere and its surrounding at SZA ~ 70° in 3 sectors of MSE coordinate system are presented in table 2.

Table 2. The properties of magnetosphere and its surrounding at SZA ~ 70° in 3 sectors of MSE coordinate system

| | Sector 0°-30° | Sector 60°-120° | Sector 160°-180° |
|---|---|---|---|
| Magnetic barrier | Magnetic barrier is found in all passes. It continues in magnetosphere with increasing magnitude. In most cases magnetic field magnitude is smaller in the ionosphere than in the magnetosphere. | Magnetic barrier is frequently structured and may be less developed. | Magnetic field in magnetic barrier is weak. Magnetic barrier is relatively thin. Mass loading of magnetosheath is small and consists of $O^+$ ions |
| Plume | Plume ions are seen in magnetosheath or magnetosphere or both. Strongest plume ions flux leads to decrease of magnetic field magnitude in magnetic barrier in magnetosphere. | Almost all passes do not show plume signatures. | None |
| Magnetosheath-magnetosphere interface | The density of magnetosheath protons decreases in the magnetosheath/magnetic barrier layer adjacent to the magnetosphere. The direction of proton velocity in the magnetosphere is sufficiently parallel to the magnetic field direction. The heavy ions move nearly perpendicular to the magnetic field direction. | Magnetosheath proton flux is diluted in front of magnetosphere, higher energy protons usually diminish faster. | Velocity of magnetosheath protons does not diminish at magnetopause Density of magnetosheath protons usually diminish in layer adjacent to magnetopause |
| Magnetopause | Magnetopause as the outer boundary of magnetosphere is characterized by relatively sharp increase of $(N_{O^+}$ and $N_{O2^+})/N_p$ ratio. In some cases heavy ions density at magnetopause is accompanied by the increase of magnetic field magnitude. | Multiple crossings are frequent | Magnetic field often has minimum at magnetopause. Density gradient of $O^+$ and $O_2^+$ ions at magnetopause is higher at magnetopause than in other sectors. |
| Magnetic field maximum location | It is usually observed within magnetosphere or at the boundary between the magnetosphere and the ionosphere. | Magnetic field has maximum in the magnetosphere | Location of maximum B varies between magnetic barrier, magnetosphere and ionosphere. |
| Magnetosphere | The density of protons strongly diminishes in the magnetosphere. The number densities of $O^+$ and $O_2^+$ ions increase from magnetopause to the interface with ionosphere by the factor of $10^2$ -$10^3$. Plume ions frequently dominate. The protons in the magnetosphere are moving at a relatively small angle to the magnetic field direction. The heavy ions move nearly perpendicular to the magnetic field direction. | The structure of plasma in magnetosphere is disturbed. There are energy dispersed signatures in the magnetosphere. | Magnetospheric ions have relatively small energy and usually wide energy spectrum |
| Interface with ionosphere | The inner boundary of magnetosphere with ionosphere is determined with smaller confidence due to diffuse nature of transition of the heavy ions energy between magnetosphere and ionosphere | | |

The majority of passes in sector #1 ($\theta_E \sim 0°$-$30°$) shows the plume (Liemohn et al, 2014, Dong et al., 2015, 2017, Andrews et al., 2016) within the magnetosphere and/or in the magnetosheath. At the same time, the density of the heavy ions in the magnetosphere is quite low when the magnetic pressure is high and plume is observed. The magnetic field magnitude usually has maximum within the magnetosphere in this MSE sector of magnetosphere. The strongest flux of the plume ions within the magnetosphere are accompanied by the magnetic field depression.

Figure 13 shows the pass of MAVEN on 01.19.2016 with the calculated directions of the electric field and the velocities of the heavy ions and protons. It is seen that the heavy ions are accelerated within the magnetosphere and produce the plume in the magnetosheath. This shows a significant fraction of the magnetospheric ions which are accelerated to form the plume thus leading to depletion of the magnetospheric ions. Consequently, the plume ions flux needs to be included to the flux of the magnetospheric ions in order to make a correct estimation of the total magnetospheric ion flux in sector #1.

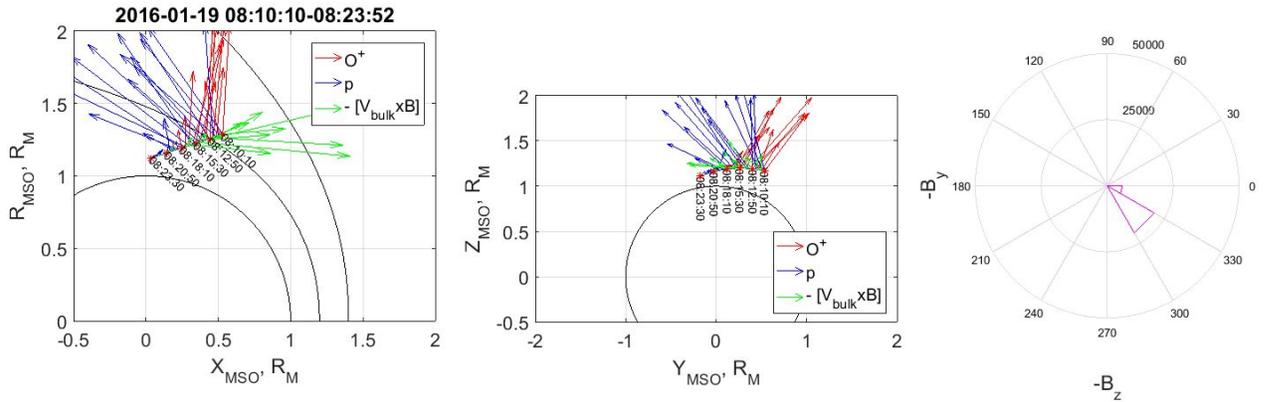

Figure 13. Electric field vectors (green), velocities of protons (blue) and O$^+$ ions (red) in cylindrical coordinates X, (X$^2$+Z$^2$)$^{1/2}$ (left), in YZ$_{MSO}$ plane (middle), and magnetosheath magnetic vectors distribution on B$_y$-B$_z$ plane (right).

The magnetic barrier in magnetic sector #1 of the magnetosphere develops in the magnetosheath and continues into the magnetosphere. A large-scale rotation of the magnetic field frequently accompanies this transition. However, it is difficult to identify the magnetic pileup boundary in passes within this MSE sector and, consequently, to make a conclusion that the magnetic pile-up boundary exists and whether plasma and magnetic boundaries coincide or not.

Sector #2 in MSE coordinate system ($\theta_E \sim 60°$-120°) shows a large variety of the magnetospheric structures, and the ion flux and the magnetic field are variable along the spacecraft path. Like in sector #1, one can often see the depletion of more energetic protons in the part of the magnetosheath adjacent to the magnetopause and within the magnetosphere. The magnetospheric flow may be in Alfvenic regime or the magnetic pressure can be higher than the plasma pressure. One can see multiple crossings of the magnetopause. These properties of the magnetospheric boundary in this sector suggest that either it is unstable or the mass loaded magnetic flux tubes are added to the magnetosphere in non-stationary regime.

The structure of the magnetosheath and the magnetic barrier in sector #3 in MSE coordinate system ($\theta_E \sim 160°$-180°) is different from other sectors. There is no depletion of more energetic protons in front of the magnetopause. However, the density of the protons drops in the magnetic barrier that is consistent with the higher magnetic field magnitude than the one in the magnetosheath. The magnetosphere thickness is smaller than in other sectors of the magnetosphere and it is located higher than the one in other sectors.

The magnetic field magnitude frequently has the maximum within the magnetosphere and decreases towards the ionosphere. It means that the pressure balance on the dayside is not provided by the ionospheric current. This makes the term "induced magnetosphere" not fully justified. This term was proposed in the earlier years of Mars and Venus investigations, mostly based on theoretical considerations. The model of Venusian magnetosphere (Vaisberg and Zelenyi, 1984; Zelenyi and Vaisberg, 1985) based on experimental data and analysis of the dayside Martian magnetosphere crossing (Vaisberg et al., 2017) gives evidence that the dayside magnetosphere is formed by the solar wind magnetic flux tubes entering magnetosphere near the subsolar region and loaded by the planetary photo-ions during tubes' convection within the dayside magnetosphere to the terminator and then into the tail.

In sector #1 the average maximum of the magnetic field in the magnetosheath-ionosphere interface region is about 25-30 nT and is higher than in sector #3 where it is about 15-20 nT. In other words, the magnetic barrier in sector #1 is stronger than in sector #3. This can be associated with higher mass loading in sector #1 by the planetary plume ions ejected from the interface region by the motional electric field.

It is interesting to compare the heights of magnetospheric boundaries with other boundaries found in near-Mars space. Duru et al., (2009) found steep gradient of electron density at about 660 km at SZA 60°- 70°. The SZA range is about the same for observations in this paper. Withers et al., (2016) found photoelectron boundary median height of 620 km and magnetic pileup boundary median height 970 km. The average location of the photoelectron boundary found by Garnier et al., (2017) at SZA ~70° is about 600 km. Our results can be compared with the electron boundaries locations of Duru et al., (2009) and Garnier et al., (2017), both obtained at SZA favorably corresponding to the location of observations used in this paper. The average height of the boundary between the ionosphere and the magnetosphere is 530 km. Thus we can conclude that this boundary is most probably the planetary electron boundary.

In sector #1 the average maximum of the magnetic field in the magnetosheath-ionosphere interface region is about 25-30 nT which is higher than in sector #3 where it is about 15-20 nT. In other words, the magnetic barrier in sector #1 is stronger than in sector #3. This can be associated with higher mass loading in the sector #1 by the planetary plume ions ejected from interface region by the motional electric field.

The magnetic field magnitude often has maximum within the magnetosphere and decreases while the spacecraft moves to the ionosphere where it is smaller than in the magnetosphere. This makes the term induced magnetosphere not well justified. The magnetosphere forms due to mass-loading of the magnetic flux tubes while they drift within the upper atmosphere of Mars. In essence, this is mass-loaded magnetosphere or accretion magnetosphere.

The estimated integral median value of calculated planetary ions ($O^+$ and $O_2^+$) flux through 1 cm of the magnetosphere at SZA ~70° corresponds to average flux density ~ 3.8 x 10^6 $cm^{-2}$ $c^{-1}$. The flux density should increase during the convection of the magnetic flux tubes to the terminator and then to the tail. The average flux of $O^+$ ions in the boundary layer in the Martian tail is ~ 2x10^5 $cm^{-2}$ $c^{-1}$ (e.g. Dubinin et al., 2017), to which magnetospheric flow is the source. As the width of the boundary layer in the tail is larger than the thickness of the dayside magnetosphere, and the calculated heavy ions flux in the dayside magnetosphere includes two ion species we can conclude that these two numbers are not contradictory.

## 5 Conclusion

The magnetosphere as specific domain between the magnetosheath flow and the ionosphere was identified in all 44 crossings of the dayside interaction region of the solar wind plasma with the Martian atmosphere at the solar zenith angle of ~ 70° in the Northern part of the planet. The magnetopause is characterized by the stepwise increase of the number densities of the planetary ions $O^+$ and $O_2^+$ to protons ratio. The number densities of $O^+$ and $O_2^+$ ions increase by 2-3 orders of magnitude across the thickness of magnetosphere which is ~200 km on average. The energy of heavy ions decreases with the decrease of the altitude until it approximately equalizes to the energy of the ionospheric ions at the interface with the ionosphere. The number densities of the planetary ions and their height profiles indicate their origin as the pickup ions inside the magnetosheath. The magnetic flux tubes enter the magnetosphere in the subsolar region and accumulate the photo-ions on their convection to the terminator.

The structure of the magnetic barrier and the magnetosphere significantly varies with the location in MSE coordinates that is defined by the motional electric field direction. The direction of this electric field relative to the mass-loaded magnetic flux tubes convection direction determines the structure of magnetosphere. There are varieties in typical magnetosheath and magnetic barrier structures observed in different sectors of MSE coordinate system. In sector #1 magnetic barrier starts to form in the magnetosheath well outside the magnetosphere and the average maximum of the magnetic field magnitude reaches the values of 25-30 nT which are higher than at other magnetic latitudes; in sector # 2 magnetic barrier is frequently quite structured and may be less developed; in sector # 3 magnetic barrier starts to form in the magnetosheath and the average maximum of the magnetic field magnitude reaches the values of 15-20 nT which are lower than in sector # 1.

**Acknowledgement**

The MAVEN project is supported by NASA through the Mars Exploration Program. MAVEN data are publicly available through the Planetary Data System. Authors O.V., L.Z., V.E. and S.S. wish to acknowledge support from the Russian Science Foundation by grant 16-42-01103. Author E. D. wishes to acknowledge DFG for supporting this work by grant PA 525/14-1 and support from DLR by grant 50QM1703. Authors are grateful to the team of the Centre de Données de la Physique des Plasmas (CDPP) for giving access to experimental data with the use of the AMDA data processing system, whose development was supported by CNRS, CNES, Observatoire de Paris and Université Paul Sabatier, Toulouse.